\documentclass[prl,twocolumn,amsmath,amssymb,groupedaddress,aps]{revtex4}

\newcommand{\xv}{{\bf x}}
\newcommand{\kv}{{\bf k}}
\newcommand{\eps}{{\varepsilon}}

\newcommand{\qv}{{\bf Q}}
\newcommand{\Qv}{{\bf Q}}
\newcommand{\jv}{{\bf j}}

\newcommand{\oh}{{\frac{1}{2}}}

\newcommand{\cH}{{\mathcal H}}

\def\rf#1{(\ref{#1})}

\newcommand{\bk}{{\bf k}}

\newcommand{\bQ}{{\bf Q}}

\newcommand{\zh}{\hat{z}}

\newcommand{\ch}{\hat{c}}

\newcommand{\Psih}{\hat{\Psi}}

\newcommand{\be}{\begin{equation}}
\newcommand{\ee}{\end{equation}}
\newcommand{\bea}{\begin{eqnarray}}
\newcommand{\eea}{\end{eqnarray}}
\newcommand{\bse}{\begin{subequations}}
\newcommand{\ese}{\end{subequations}}

\input{epsf}
\usepackage[dvips]{graphicx}

\begin{document}

\title{Quantum liquid crystals in imbalanced Fermi gas: fluctuations
  and fractional vortices in Larkin-Ovchinnikov states}
\author{Leo Radzihovsky} 
\author{Ashvin Vishwanath}
\affiliation{Department of Physics, University of Colorado,
   Boulder, CO 80309}
\affiliation{Department of Physics, University of California,
   Berkeley, CA 94703}

\date{\today}

\begin{abstract}
  We develop a low-energy model of a unidirectional Larkin-Ovchinnikov
  (LO) state. Because the underlying rotational and translational
  symmetries are broken spontaneously, this gapless superfluid is a
  smectic liquid crystal, that exhibits fluctuations that are
  qualitatively stronger than in a conventional superfluid, thus
  requiring a fully nonlinear description of its Goldstone
  modes. Consequently, at nonzero temperature the LO superfluid is an
  algebraic phase even in 3d. It exhibits half-integer
  vortex--dislocation defects, whose unbinding leads to transitions to
  a superfluid nematic and other phases. In 2d at nonzero temperature,
  the LO state is always unstable to a nematic superfluid.  We expect
  this superfluid liquid-crystal phenomenology to be realizable in
  imbalanced resonant Fermi gases trapped isotropically.
\end{abstract}
\maketitle

The tunability of interactions through Feshbach resonances has led to a
realization of an s-wave paired superfluidity and BCS-BEC
crossover\cite{Regal_swave,Zwierlein_swave,Kinast_swave}, as well as
promises of more exotic states such as gapless p-wave\cite{pwave} and
periodic Fulde-Ferrell-Larkin-Ovchinnikov (FFLO)\cite{FF,LO}
superfluidity in strongly correlated degenerate alkali gases. The
latter enigmatic state has been thoroughly explored within a BCS
mean-field studies\cite{MachidaNakanishiLO,BurkhardtRainerLO,MatsuoLO}
and is expected to be realizable in a population-imbalanced
(polarized) Feshbach resonant Fermi
gas\cite{MizushimaLO,SRimbalanced}. While recent
experiments\cite{Zwierlein06imbalanced,Partridge06imbalanced}, have
confirmed much of the predicted phenomenology of phase
separation\cite{Combescot,Bedaque,SRimbalanced} in such systems, the
FFLO states have so far eluded definitive observation.

The simplest mean-field treatments\cite{FF,LO,SRimbalanced} find that
the FFLO type states are quite fragile, confined to a narrow range of
polarization on the BCS side. However, motivated by earlier studies
\cite{MachidaNakanishiLO,BurkhardtRainerLO} and based on the finding
of a negative domain-wall energy in an otherwise uniform singlet BCS
superfluid\cite{MatsuoLO,YoshidaYipLO}, it has recently been argued
that a more general periodic superfluid state that includes a larger
set of collinear wavevectors may be significantly more stable. Much
like a type-II superconductor undergoes a continuous transition into a
vortex state at a lower-critical field $H_{c1}$, which is
significantly below the thermodynamic field, here too, a Zeeman-field
driven domain-wall nucleation (with the density increasing above the
lower critical $h_{c1}$ field) allows a continuous mechanism for a
transition from a singlet paired superfluid to a LO-like periodic
state\cite{MachidaNakanishiLO,BurkhardtRainerLO,MatsuoLO,YoshidaYipLO}.

\begin{figure}[bth]
\vspace{2.5cm}
\centering
\setlength{\unitlength}{1mm}
\begin{picture}(40,45)(0,0)
\put(-18,2){\begin{picture}(0,0)(0,0)
\includegraphics{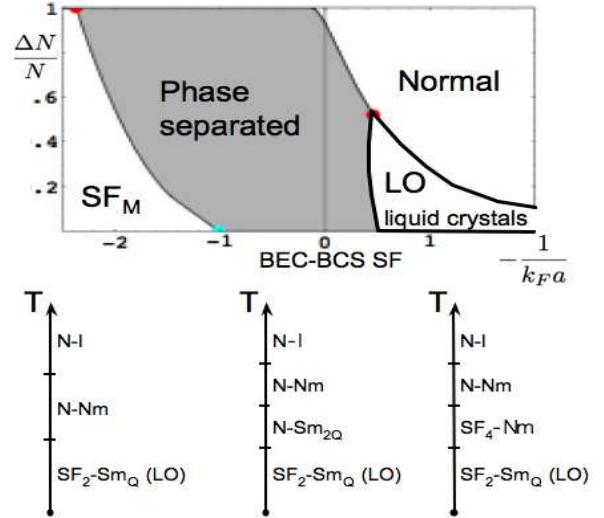}
\end{picture}}
%
\end{picture}
\vspace{-.5cm}
\caption{Polarization $\Delta N/N$ vs. $1/(k_F a)$ schematic phase
  diagram, showing LO liquid crystal phases replacing phase-separated
  (PS) regime\cite{SRimbalanced}. Possible 3d transition scenarios as
  a function of temperature to the normal-nematic (N-Nm),
  normal-isotropic (N-I), normal-smectic (N-Sm$_{2Q}$), and ``charge''-4
  superfluid-nematic (SF$_4$-Nm) phases are illustrated in the lower
  panel.}
\label{phasediagram}
\end{figure}

In this scenario the SF-LO transition is of a
commensurate-incommensurate type
as can be explicitly shown in one dimension
(1d)\cite{MachidaNakanishiLO,Yang1D}. The imposed species imbalance
(excess of the majority fermionic atoms) can be continuously
accommodated by the subgap states localized on the self-consistently
induced domain-walls, with this picture resembling the doping of
polyacetylene\cite{SuSchriefferHeeger}. Such LO state can also be
thought of as a periodically ordered {\em micro}-phase separation
between the normal and BCS states, that thus naturally replaces the
{\em macro}-phase separation ubiquitously found in the BEC-BCS
detuning-polarization phase diagram\cite{SRimbalanced}
(Fig.\ref{phasediagram}).

With this motivation in mind, here we report on our study that is
complementary to these microscopic mean-field investigations.  Namely,
assuming that the LO state is indeed energetically favorable over a
region of a phase diagram, we explore its stability to low-energy
fluctuations and the resulting phenomenology.

We demonstrate that the low-energy model of the LO state is that of
two coupled smectics, whose moduli we derive from the BCS theory.
Thus we show that a resonant imbalanced Fermi gas, confined to an
isotropic trap is a natural realization of a quantum (superfluid)
liquid crystal, that unlike the solid state
analogs\cite{Fogler,KivelsonFradkinEmery,Berg,RadzihovskyDorsey} is
not plagued by the underlying lattice potential that explicitly breaks
continuous spatial symmetries.

We find that while it is stable to quantum fluctuations, in 3d a
long-ranged LO order is marginally unstable at any nonzero $T$.  The
resulting superfluid state is an {\em algebraic} phase, characterized
by universal {\em quasi}-Bragg peaks and correlations that admit an
asymptotically exact description\cite{GP}. In contrast, crystalline LO
phases with multiple noncolinear ordering wavevectors are stable
against thermal fluctuations.

As with treatments of the LO state where long range order is
assumed\cite{BergPRL,Agterberg,Berg}, we also find an unusual
topological excitation -- a half vortex bound to a half dislocation --
in addition to integer vortices and dislocations, in the algebraic LO
phase.  Because our conclusions are based on general symmetry
principles, supported by detailed calculations, they are generic and
robust to variations in microscopic details.

In 2d and nonzero $T$, the state is also an {\em algebraic} phase,
exhibiting universal power-law phonon correlations, controlled by a
nontrivial exactly calculable\cite{GW} fixed point. It displays
short-range positional order with Lorentzian structure function peaks,
and is thus unstable to proliferation of dislocations. The resulting
state is either a ``charge''-$4$ (four-fermion) superfluid or a
non-superfluid nematic, depending on the relative energetics of
aforementioned integer and half-integer vortex-dislocation defects.
The latter normal nematic state is a (complementarily
described\cite{RadzihovskyDorsey}) deformed Fermi surface
state\cite{Oganesyan,Sedrakian}.

%

{\em MODEL:} We now present the highlights of our calculations. We
begin with a Ginzburg-Landau theory that captures the system's
tendency to order into a finite wavevector paired state, with a
preferred magnitude $Q_0$ and a spontaneously chosen direction.  The
corresponding free-energy density
\begin{equation}
\cH = J\left[|\nabla^2\Delta|^2 - 2Q_0^2|\nabla\Delta|^2\right] +
r|\Delta|^2 + \frac{v_1}{2}|\Delta|^4 + \frac{v_2}{2}\ \jv^2 + \ldots.
\label{cHsm} 
\end{equation}
can be derived from a microscopic BCS
model\cite{LO,SRimbalanced,LRnotes} near the upper-critical chemical
potential difference, $h_{c2}$, with 
\begin{eqnarray}
J&\approx&\frac{0.61n}{\epsilon_FQ_0^4},\ 
Q_0\approx\frac{1.81\Delta_{BCS}}{\hbar v_F},\ 
r\approx\frac{3n}{4\epsilon_F}\ln\left[\frac{9h}{4h_{c2}}\right],\nonumber\\
h_{c2}&\approx&\frac{3}{4}\Delta_{BCS},\ 
v_1\approx\frac{3n}{4\epsilon_F\Delta_{BCS}^2},\ 
v_2\approx\frac{1.83n m^2}{\epsilon_F\Delta_{BCS}^2Q_0^2},\ \ \ \ 
\label{moduli}
\end{eqnarray}
that can be more generally taken as phenomenological parameters to be
determined experimentally. $n$, $\epsilon_F$, $v_F$, $\Delta_{BCS}$,
and $m$ are the atomic density, Fermi energy and velocity, BCS ($h=0$)
gap and atomic mass, respectively. Near the lower-critical Zeeman
field, $h_{c1}$, $Q_0(h)$ is expected to vanish with the species
imbalance, as the system continuously transitions into a uniform
singlet
superfluid\cite{MachidaNakanishiLO,BurkhardtRainerLO,MatsuoLO}, with
this and other moduli's dependences derivable via fluctuating
domain-walls methods\cite{LRnotes}.  Above, $\jv$ is the supercurrent
and the last term is crucial for getting a nonzero transverse (to
$\Qv$) superfluid stiffness in the LO state.  From the first two terms
it is clear that the dominant instability and fluctuations are near a
finite wavevector of magnitude $Q_0$. Thus, for $h<h_{c2}$,
$r<JQ_0^4=0.61n/\eps_F$ and the system develops a pairing order
parameter $\Delta(\xv)=\sum_{\Qv_n}\Delta_{\Qv_n} e^{i\Qv_n\cdot\xv}$.

As with other crystallization problems, the choice of the set of
$\Qv_n$'s is determined by the details of interactions and will not be
addressed here. Motivated by LO findings\cite{LO}, we focus on the
unidirectional order characterized by a collinear set of
$\Qv_n$'s. These fall into two, LO and FF universality classes.  The
LO (FF) states are characterized by breaking (preserving)
translational and preserving (breaking) time-reversal symmetries.
Low-energy properties of such states can be well captured with a
single $\pm\Qv$ pair (LO) and a single $\Qv$ wavevector (FF)
approximations.

Because it is the periodic LO state that is expected to be most
stable\cite{LO,MachidaNakanishiLO,BurkhardtRainerLO,MatsuoLO,YoshidaYipLO},
we focus on this more interesting case and only comment in passing on
the homogeneous FF state. Within the LO approximation the pairing
function is given by $\Delta_{LO}(\xv) =\Delta_+(\xv) e^{i\qv\cdot\xv}
+ \Delta_-(\xv) e^{-i\qv\cdot\xv}$, where $\Delta_\pm = \Delta_Q
e^{i\theta_\pm(\xv)}$ are the leading complex order parameters, whose
amplitudes deep in the ordered LO state can be taken to be equal and
constant, $\Delta_Q^2\approx c\Delta_{BCS}^2\ln(h_{c2}/h)$, thereby
focusing on the two Goldstone modes $\theta_\pm(\xv)$.  A slightly
rearranged form of the LO order parameter $\Delta_{LO}(\xv)$ clarifies
its physical interpretation
\begin{eqnarray}
\hspace{-1cm}
\Delta_{LO}(\xv)
&=&2\Delta_Qe^{i\theta_{sc}(\xv)}
\cos\big(\qv\cdot\xv + \theta_{sm}(\xv)\big),
\label{LOop}
\end{eqnarray}
showing that it is a product of a superfluid order parameter and a
unidirectional, spontaneously oriented (along $\Qv$) Cooper-pair
density wave, i.e., simultaneously exhibiting the ODLRO and smectic
order.  The low-energy properties are respectively characterized by
two Goldstone modes, the superconducting phase
$\theta_{sc}\equiv\oh(\theta_+ + \theta_-)$ and the phonon
displacement $u=\theta_{sm}/Q\equiv\oh(\theta_+ - \theta_-)/Q$. In
contrast, the uniform FF state is characterized by a single $\Delta_Q$
amplitude and a Goldstone mode $\theta_Q$.

Substituting $\Delta_{LO}(x)$ into $\cH$ we obtain a Hamiltonian
density for the bosonic Goldstone modes of a generic LO state:
\begin{eqnarray}
\cH_{LO} &=&\sum_{\alpha=\pm}\left[\frac{K}{4}(\nabla^2 u_\alpha)^2 + 
\frac{B}{4}\bigg(\partial u_\alpha + \frac{1}{2}(\nabla u_\alpha)^2\bigg)^2
\right]\nonumber\\
&&+ \frac{\gamma}{2}\left(\nabla u_+ - \nabla u_-\right)^2,\label{H_LO}\\
&\approx&\frac{K}{2}(\nabla^2 u)^2 + 
\frac{B}{2}\bigg(\partial u + \frac{1}{2}(\nabla u)^2\bigg)^2
+\frac{\rho_s^i}{2}(\nabla_i\theta_{sc})^2,\nonumber
\end{eqnarray}
where we dropped constant and fast oscillating parts, chose $\Qv =
Q_0\zh$, and defined phonon fields $u_\pm = \pm\theta_\pm/Q_0$ and the
bend ($K=4J Q_0^2\Delta_Q^2\approx 2.4 n
\Delta_Q^2/(\epsilon_FQ_0^2)$) and compressional ($B=16J
Q_0^4\Delta_Q^2\approx 9.8 n \Delta_Q^2/\epsilon_F$) elastic moduli.

This form (valid beyond above weak-coupling microscopic derivation) is
familiar from studies of conventional smectic liquid
crystals\cite{deGennes}, with rotational invariance encoded in two
ways. Firstly, for a vanishing $\gamma$ the gradient elasticity in
$u_\pm$ (and $u$) only appears {\em along} $\qv$, namely
$\partial\equiv\hat{\qv}\cdot\nabla$ (compression), with elasticity
transverse to $\qv$ of a ``softer'' Laplacian (curvature)
type. Secondly, the elastic energy is an expansion in a
rotationally-invariant strain tensor combination $u_{QQ}^\pm=\partial
u_\pm +\oh(\nabla u_\pm)^2$, whose nonlinearities in $u_\pm$ ensure
that it is fully rotationally invariant even for large reorientations
$Q_0\zh\rightarrow\Qv$ of the LO ground state.

A nonzero $\gamma\equiv v_2Q_0^2\Delta_Q^4/m^2\approx 1.8
n\Delta_Q^4/(\epsilon_F\Delta_{BCS}^2)$ coupling (minimized by a
vanishing supercurrent $\nabla\theta_+ + \nabla\theta_-$) removes the
two independent rotational symmetries, orientationally locking the two
smectics. This leads to the superconducting phase combination,
$\theta_{sc}=\oh(\theta_+ + \theta_-)$ to be of a conventional XY (as
opposed to ``soft'' smectic) gradient type. It is characterized by
parallel ($\rho_s^{i=\parallel}\approx B/Q_0^2$) and transverse
($\rho_s^{i=\perp}\equiv 4\gamma/Q_0^2$) superfluid stiffnesses,
appearing in the second form of $\cH_{LO}$, equivalent to the first
form at low energies of interest to us.

Physically, $\rho_s^\parallel$ and $\rho_s^\perp$ are superfluid
stiffnesses for the supercurrent $\jv = (\jv_++\jv_-)/2$ produced by
the imbalance in the left ($\jv_-$) and right ($\jv_-$) supercurrent
magnitudes and directions, respectively.  We thus find that the LO
state is a highly anisotropic superfluid, with
\begin{equation}
\rho_s^\perp/\rho_s^\parallel=
\frac{3}{4}\left(\Delta_Q/\Delta_{BCS}\right)^2
\approx\ln(h_{c2}/h)\ll 1,
\label{ratio}
\end{equation}
a ratio that vanishes for $h\rightarrow h_{c2}^-$.   We find the FF
state to be even more exotic, characterized by an identically
vanishing transverse superfluid stiffness, a reflection of the
rotational invariance of the spontaneous current to an
energy-equivalent ground state.

{\em FLUCTUATIONS:} The thermodynamics can be obtained in a standard
way through a coherent path-integral. Although there are nontrivial
issues of the interplay between the fermionic quasi-particles and the
Goldstone modes,
we can unambiguously show that at $T = 0$ the superfluid and smectic
orders (and thus the LO state) are stable to quantum fluctuations in
$d > 1$\cite{LRnotes}.  

In contrast, for $T>0$, $\theta_{sc}$ ($\theta_{sm}$) fluctuations
diverge and ODLRO (smectic order) is destroyed for $d \leq 2$ ($d \leq
3$).  Consequently, we find that the LO state is unstable to thermal
fluctuations, displaying quasi-Bragg (Lorentzian) peaks in 3d (2d) in
its structure function. Thus in both cases the LO order parameter,
\rf{LOop} vanishes and the state is qualitatively distinct from its
mean-field form, at low $T$ characterized by a ``charge''-4 superfluid
order parameter
$\Delta_{sc}^{(4)}\sim\Delta^2\approx\oh\Delta_Q^2e^{i2\theta_{sc}}$.

Furthermore, in the presence of these divergent thermal fluctuations
phonon nonlinearities in $\cH_{LO}$, Eq.\rf{H_LO} become
important. They qualitatively modify correlations on scales longer
than $\xi_{NL}\sim \big[K^{3/2}/(B^{1/2}T)]^{1/(3-d)}\sim
k_F^{-1}[\Delta_Q^2\epsilon_F/(\Delta_{BCS}^3 T)]^{1/(3-d)}$ (on
shorter scales the harmonic description above applies), giving
universal power-laws, e.g., $\langle u(z,x)u(0,0)\rangle^{1/2}\sim
{\text{ Max}}\left[x^{\alpha}, z^{\beta}\right]$, controlled by a
nontrivial low $T$ (order $3-d$) fixed point\cite{GP}, that has an
exact description in 2d with $\alpha=1/2,\beta=1/3$\cite{GW}.  In 3d,
$\xi_{NL}\sim e^{c K^{3/2}/(B^{1/2}T)}$ and phonon correlations grow
as a universal power of a logarithm, a result that is asymptotically
exact. These elastic results of course only hold as long as
dislocations remain bound or on scales shorter than the dislocation
unbinding scale.

{\em DEFECTS:} We now turn to the discussion of topological defects
and corresponding phases accessible by their unbinding. With two
compact Goldstone modes $\theta_{sc}, u$ (equivalently,
$\theta_\pm=\pm 2\pi u_{\pm}/a$), defects are labeled by vortex and
dislocation charges $(2\pi n_{v}, a n_{d})$.  Ordinary vortex,
$(2\pi,0)$ and dislocation $(0,a)$ are clearly allowed, and in terms
of the two smectic displacements these respectively correspond to the
opposite and same signs of integer dislocations in $u_\pm$. When
proliferated they destroy the superfluid phase coherence and smectic
periodic order, respectively, and either one is clearly sufficient to
suppress the conventional LO order, $\Delta_{LO}$, \rf{LOop}.

However, because a sign change in $\Delta_{LO}$ due to a
$a/2$-dislocation in $u$ can be compensated by a $\pi$-vortex in
$\theta_{sc}$ (thereby preserving a single-valuedness of
$\Delta_{LO}$) $1/2$-charge defects in $\theta_{sc}$ and $u$ are also
allowed, but are confined into $(\pm\pi,\pm a/2)$
pairs\cite{BergPRL,Agterberg,Berg}.  In terms of the two coupled
smectic displacement fields, $u_+,u_-$ these correspond to an integer
dislocation in one and no dislocation in the other.

{\em TRANSITIONS:} There are thus many paths of continuous transitions
out of the LO (SF$_2$-Sm$_Q$) state. One is through an unbinding of
ordinary integer $(0,a)$ dislocations in $u$. This melts the smectic
order in favor of a nematic, but retains a superfluid order, thereby
transforming the LO state to a nematic ``charge''-$4$ superfluid
(SF$_4$-Nm).  Another path, is by unbinding integer $(2\pi,0)$
vortices in $\theta_{sc}$. This destroys the superfluid order and
converts the smectic positional order $Q$ to $2Q$
(N-Sm$_{2Q}$). Finally, a third route out of the LO superfluid is
through a direct proliferation of $(\pi,\pm a/2)$ fractional
vortex-dislocation pairs, that destroy both smectic and superfluid
orders, inducing a transition to a normal (nonsuperfluid) nematic
(N-Nm). For 3d these possibilities, determined by the relative
energetics of these different types of defects are illustrated in
Fig.\ref{phasediagram}. In 2d, the dislocation energy is finite and LO
state is {\em necessarily} destabilized by thermal fluctuations to a
``charge''-$4$ superfluid nematic, SF$_4$-Nm.  Upon rotation the resulting
nematic superfluid will display $\pi$ vortices ($\oint{\bf v}\cdot{\bf
  dl}=h/4m$), that (because of the nematic order) we expect to form a
uniaxially distorted lattice. We note that this rich
fluctuations-driven phase behavior contrasts sharply with a direct
LO-N transition (described by $U(1)\times U(1)$ Landau theory
$H_{mft}=r(|\Delta_+|^2+|\Delta_-|^2) + \lambda_1
(|\Delta_+|^4+|\Delta_-|^4) +\lambda_2|\Delta_+|^2|\Delta_-|^2$) found
in mean-field theory.

{\em FERMIONS:} We now turn to a discussion of the fermionic sector
that we have so far ignored. Near $h_{c2}$ a single harmonic ($Q$ for
FF and $\pm Q$ for LO states) approximation is sufficient. Unlike the
simpler FF case (that can be diagonalized exactly with a two-component
Nambu spinor\cite{SRimbalanced}), the LO state involves a
three-component spinor $\Psih_\bk
\equiv(\ch_{-\bk+\bQ\uparrow},\ch_{\bk\downarrow}^{\dagger},\ch_{-\bk-\bQ\uparrow})$.
Neglecting sparse off-resonant coupling between $\kv$ and $\kv+2\Qv$
Cooper-pairs and noting that only two of the three components in
$\Psih_\bk$ are resonant at any one $\bk$, the approximate spectrum is
given by $E_{\bk,\sigma,\pm\Qv} =(\varepsilon_k^2 +\Delta_\bQ^2)^{1/2}
- \sigma\big(h \pm \frac{\bk \cdot \bQ}{2m}\big)$, with $\sigma=\pm1$,
$\varepsilon_k = \frac{k^2}{2m} - \mu + \frac{Q^2}{8m}$, $\mu =
\oh(\mu_\uparrow + \mu_\downarrow)$, and $h=\oh(\mu_\uparrow -
\mu_\downarrow)$. The regions of $\kv$ where $E_{\bk,\sigma,\pm\Qv}$
is negative corresponds to a Fermi sea (rather than the usual vacuum)
of Bogoluibov quasi-particles in the BCS ``vacuum'', and therefore
leads to a Fermi surface of gapless fermionic excitations. These are
nothing but the unpaired fraction of the majority atoms. From the
spectrum above it is clear that two distinct LO states are
possible. One, LO1 exhibits a single (majority) species, fully
polarized Fermi surface pockets. The other, LO2 is characterized by
both majority and minority fermion flavor Fermi surface pockets. The
Fermi surface volume difference is proportional to the species
imbalance and the anisotropy encodes the long-range orientational
order of the LO state.  Because the polarization is conserved and
Goldstone mode fluctuations are finite at zero temperature, we expect
the LO1-LO2 quantum transition to be of a simple band filling type.

In the complementary regime near $h_{c1}$, the excess majority atoms
occupy additional states localized on
domain-walls\cite{MachidaNakanishiLO,BurkhardtRainerLO,MatsuoLO,YoshidaYipLO}.
Because the atoms can freely move along and tunnel between adjacent
domain-walls, near $h_{c1}$ they exhibit a ``metallic'', but highly
anisotropic dispersion. The resulting fermionic spectrum is consistent
with that near $h_{c2}$, $E_{\bk,\sigma,\pm\Qv}$.

The interactions between the Goldstone modes and unpaired majority
fermionic atom, $\psi$ must now be included and are given by
\begin{eqnarray}
\hspace{-1cm}
\cH_{j_s,j}&\sim&i\nabla\theta\cdot\psi^{\dagger}\nabla\psi + h.c.,\ \
\ \cH_{j_s,n}\sim(\nabla\theta)^2\psi^\dagger\psi,\ \ \\
\cH_{a-p}&\sim&\big[\partial_z u + \frac{1}{2}(\nabla u)^2\big]
\psi^\dagger\psi + i\nabla u\cdot\psi^\dagger\nabla\psi + h.c.\nonumber
\end{eqnarray}
As with other analogous problems\cite{Oganesyan}, we expect these to
lead to Landau-like damping of the Goldstone modes $\theta_{sc},u$,
and a finite fermionic quasi-particle lifetime.  We leave the study of
these and other affects on the properties of the LO states to the
future.

{\em TRAP EFFECTS:} Since near $h_{c2}$ the LO period
$\lambda_Q=2\pi/Q_0$ \rf{moduli} is bounded by the coherence length
(that near unitarity can be as short as $\sim R/N^{1/3}$, where $R$ is
the trapped condensate radius and $N$ is the total number of atoms),
and thus $\ll R$, in this regime the trap can be treated via a local
density approximation (LDA). For $\lambda_Q\ll R$, LDA predicts weak
pinning of the LO smectic, that can be estimated via finite size
scaling, with trap size $R$ cutting off $\langle
u^2\rangle\sim\eta\log(R/\lambda_Q)$, leading to
$\langle\Delta_{LO}\rangle\sim (\lambda_Q/R)^\eta\ll 1$ that no longer
truly vanishes, but is still strongly suppressed.  We thus expect the
predicted strong fluctuations effects to be experimentally
accessible. We note, for example, that Kosterlitz-Thouless phase
fluctuation physics has been reported in 2d trapped
superfluids\cite{Hadzibabic}, despite the finite trap size.  However,
a more detailed analysis of the trap effects is necessary,
particularly near $h_{c1}$, for a quantitative comparison with
experiments.

To summarize, we studied fluctuation phenomena in a LO state, expected
to be realizable in imbalanced resonant Fermi gases. The LO state is a
superfluid smectic liquid crystal, whose elastic moduli and superfluid
stiffness we derived near $h_{c2}$. It is extremely sensitive to
thermal fluctuations that destroy its long-range positional order even
in 3d, replacing it by an algebraic phase, that exhibits vortex
fractionalization, where the basic superfluid vortex is half the
strength of a vortex in a regular paired condensate. This should be
observable via a doubling of a vortex density in a rotated state.
Also under rotation, the high superfluid anisotropy \rf{ratio} leads
to an imbalance-tunable strongly anisotropic vortex core and a
lattice highly stretched along $\Qv$. Bragg peaks in the
time-of-flight images can distinguish the periodic SF$_2$-Sm$_Q$
(superfluid smectic) state from the homogeneous SF$_4$-Nm (superfluid
nematic), which are in turn distinguished from the N-Sm$_{2Q}$ and
N-Nm (normal smectic and nematic) by their superfluid properties,
periodicity, collective modes, quantized vortices, and condensate
peaks. Thermodynamic signatures will identify corresponding phase
transitions.

We acknowledge support from the NSF DMR-0321848 (LR), DMR-0645691
(AV), the Miller and CU Faculty Fellowships (LR), and thank
S. Kivelson for helpful comments on the manuscript. LR thanks Berkeley
Physics Department for its hospitality.

\end{document}